\documentclass[pre,a4paper,superscriptaddress,twocolumn,showpacs,amsmath,amssymb,floatfix,nofootinbib]{revtex4}
\pdfoutput=1

\usepackage{graphicx}
\usepackage{amssymb}
\usepackage{float}

\usepackage{color}

\begin{document}

\title{Random matrix model of Kolmogorov-Zakharov turbulence}

\author{Klaus M.~Frahm}
\affiliation{\mbox{Laboratoire de Physique Th\'eorique, IRSAMC, 
Universit\'e de Toulouse, CNRS, UPS, 31062 Toulouse, France}}
\author{Dima L.~Shepelyansky}
%\homepage[]{http://www.quantware.ups-tlse.fr}
%\email[Corresponding author: ]{dima@irsamc.ups-tlse.fr}
\affiliation{\mbox{Laboratoire de Physique Th\'eorique, IRSAMC, 
Universit\'e de Toulouse, CNRS, UPS, 31062 Toulouse, France}}

%\date{today}
\date{January 21, 2024}

\begin{abstract}
We introduce and study a random matrix model of Kolmogorov-Zakharov 
turbulence in a nonlinear purely dynamical finite size system with many 
degrees of freedom. For the case of a direct cascade the energy and norm 
pumping takes place at low energy scales with absorption at high energies. 
For a pumping strength above a certain chaos border 
a global chaotic attractor appears with a stationary energy flow through 
a Hamiltonian inertial energy interval. In this regime, 
the steady-state norm distribution is 
described by an algebraic decay with an exponent in agreement with 
the Kolmogorov-Zakharov theory. Below the chaos border the system is located 
in the quasi-integrable regime similar to the Kolmogorov-Arnold-Moser theory 
and the turbulence is suppressed. For the inverse cascade the system rapidly 
enters a strongly nonlinear regime where the weak turbulence description is 
invalid. We argue that such a dynamical turbulence is generic showing that 
it is present in other lattice models with disorder and Anderson 
localization. We point out that such dynamical models can be realized in 
multimode optical fibers.
\end{abstract}

%\pacs{95.35.+d, 32.80.Rm, 05.45.Mt,  95.10.Fh}

%Dark matter, 95.35.+d
%Rydberg states atoms, 32.80.Rm
%Anderson localization disordered solids, 71.23.An
%atomic photoionization, 32.80.Fb
%chaos astronomy 95.10.Fh
%05.45.Mt 	Quantum chaos; semiclassical methods

%05.45.-a Nonlinear dynamics and chaos
%67.85.Hj 	Bose-Einstein condensates in optical potentials
%47.27.-i 	Turbulent flows
%72.15.Rn 	Localization effects (Anderson or weak localization) 
%
%47.35.-i 	Hydrodynamic waves
%47.35.Bb 	Gravity waves 
%89.75.-k 	Complex systems 

\maketitle

\section{Introduction} 
\label{sec1}
The Kolmogorov concept of turbulence \cite{kolm41,obukhov}
assumes emergence of  energy flow through an inertial interval 
from large spacial scales, with
 an external pumping,
to small scales where energy is absorbed
by dissipation. The scaling arguments 
lead to appearance of 
%%a polynomial energy distribution
%% ``polynomial'' => only positive exponents ?!
%% I think ``power law'' is better/correct. 
a power law energy distribution
over wave modes for hydrodynamics turbulence
\cite{kolm41,obukhov}.
This concept was generalized and extended 
to weak wave turbulence, based on diagrammatic
techniques and the kinetic equation, 
%indeed showing  the emergence of polynomial
indeed showing  the emergence of power law 
distributions for various types of 
weakly interacting nonlinear waves
\cite{filonenko,zakharovbook,nazarenkobook,luka,galtier}.
This theory became known as Kolmogorov-Zakharov (KZ) turbulence
(or spectra) \cite{nazarenkobook,luka,galtier}.
In spite of various successful confirmations of this theory
in experiments and numerical simulations(see e.g. 
\cite{nazarenkobook,luka,galtier})
it is still based on
a fundamental hypothesis directly
stated in the seminal work of Zhakharov and Filonenko \cite{filonenko}:
{\it ``In the theory of weak turbulence nonlinearity of waves is assumed to be
small; this enables us, using the hypothesis of the random nature of 
the phase of individual waves, to obtain the kinetic equation 
for the mean square of the wave amplitudes''}.
Nevertheless, the dynamical equations
for waves do not involve Random Phase Approximation (RPA)
and hence, the validity for the whole concept 
of energy flow from large to small scales remains open.

Indeed, a flow through an inertial interval is 
Hamiltonian and it is well known that in nonlinear systems
with weak nonlinearity the Kolmogorov-Arnold-Moser (KAM) theory
guarantees that the main part of the system phase space
remains integrable and non-chaotic in the limit of 
very weak nonlinearity (see e.g. \cite{arnold,sinai};
note that KAM theory is valid in absence of exact resonances
as discussed below).
More physical analysis
of nonlinear dynamical Hamiltonian systems
also shows the existence of a chaos border
below which the phase space contains mainly an integrable dynamics
opposite to the turbulent one \cite{chirikov,lichtenberg}.
%Also a simple observation of a sea surface clearly shows
%that a weak wind is not able to generate turbulence.

Therefore to understand better the 
fundamental aspects of KZ turbulence (KZT), 
we introduce and study here a new Random Matrix Model (RMM) of KZT 
described only by dynamical equations of motion. 
This model is an extension of 
the Nonlinear Random Matrix model (NLIRM) introduced recently 
in \cite{rmtprl} and which describes 
a dynamical system of linear oscillators coupled 
by a random matrix combined with a nonlinear interaction between oscillators
in the form of a quartic nonlinearity corresponding to 
four-wave interactions in nonlinear media. 
This system is Hamiltonian with two conserved integrals of motion. 

Random Matrix Theory (RMT), introduced by Wigner \cite{wigner},
describes generic spectral
properties of complex nuclei, atoms and molecules  \cite{mehta,guhr}, and
systems of quantum chaos \cite{bohigas,haake}.
In particular, RMT eigenstates are ergodic and uniformly distributed on the 
$N$-dimensional unit sphere, and the level spacing statistics
is characterized by the universal RMT distribution.

These ergodic RMT eigenstates provide nonlinear long range couplings 
between the oscillator modes in the NLIRM which leads 
for a rather weak nonlinearity (but still above a certain chaos border) 
to dynamical thermalization 
according to classical statistical mechanics \cite{landau} 
with a steady-state thermal distribution characterized by energy equipartition 
\cite{rmtprl}. 

In particular, in the NLIRM the dynamical thermalization 
appears in absence of any thermal bath. This is possible because 
due to the ergodic RMT eigenstates the NLIRM allows to avoid 
specific features of nonlinear oscillator models
which can be close to certain completely integrable systems 
which makes it difficult to achieve dynamical thermalization. 
Indeed, this happened for the seminal
Fermi-Pasta-Ulam (FPU) problem \cite{fpu}, which appeared to be close
to completely integrable soliton systems such as 
the Korteweg-De Vries equation \cite{greene},
the nonlinear Schr\"odinger equation \cite{zakharov}
or the Toda lattice \cite{toda}.

Due to these reasons we think that the NLIRM \cite{rmtprl}
can be used as a basis for dynamical modeling of KZ turbulence.
For this we extend the NLIRM by additional terms describing 
pumping with nonlinear saturation
at low energies and dissipation at
high energies. The form of such pumping is
rather standard being used in systems of
fluid mechanics \cite{landau6}
and models of random lasing (see e.g. \cite{lapteva,rlaser}).
We call this extended model RMM of KZ turbulence.

Our studies for this new model show that the direct cascade
in this system is described by an algebraic decay
of the KZ turbulent spectrum \cite{zakharovbook} with
an exponent being close to the expected theoretical value,
if the pumping strength is above a certain chaos border, 
while below this border 
the flow to high energies is suppressed
in analogy with KAM theory.
The properties of the inverse cascade are
more complex as we discuss below.

The paper is organized as follows: in section II we remind the NLIRM 
of \cite{rmtprl} and generalize it to include pumping and dissipation 
at certain energy modes. Section III provides some theoretical conclusions 
based on the KZ theory of \cite{zakharovbook} for our model. 
Section IV presents and analyzes the numerical results for three different 
cases of a direct and an inverse cascade of the RMM and also the direct 
cascade for a variant with short range oscillator couplings similar 
to the Anderson 1D model and section V provides the final discussion. 

\section{Model description}
\label{sec2}

In absence of pumping and dissipation the RMM of KTZ is
reduced to the NLIRM model studied in \cite{rmtprl}
with the time evolution described by
\begin{equation}
  i\hbar{\partial\psi_n(t) \over\partial t} =  \sum_{n'=1}^N H_{n,n'} \psi_{n'}(t) 
+   \beta \vert\psi_n(t)\vert^2\psi_n(t) \;\;\;\; .
\label{eq1}
\end{equation}
Here $H_{n,n'}$ are elements of an RMT matrix $\hat H$ of size $N$ generated
from the Gaussian Orthogonal Ensemble (GOE) \cite{mehta},
they have zero mean and 
variance $\langle H_{n,n'}^2\rangle = (1+\delta_{n,n'})/(4(N+1))$.
The averaged density of states is described by the 
the semi-circle law 
$dm/dE = \frac{2N}{\pi}\sqrt{1-E^2}$ with typical eigenvalues 
in the interval $E_m\in[-1,1]$ 
(we use dimensionless units with $\hbar=1$).
Here, $\beta$ is a dimensionless constant characterizing 
the nonlinear interaction strength in the original basis $n$.
For most energies close to the band center, 
we can consider that 
the eigenenergies $E_m$ are changing approximately
linearly with $m$ ($E_m \approx \pi(m-N/2)/(2N); \; 1 \leq m \leq N$),
however, keeping in mind that at the spectrum
boundaries the density of states drops
significantly.

We denote by $\phi_n^{(m)}$ the eigenmodes of $\hat{H}$ at eigenenergies $E_m$.
They are ergodic with a uniform distribution on the $N$-dimensional 
unit sphere \cite{mehta} for fixed $m$ and mutually orthogonal between 
different $m$. 
The time evolution of the system wave function can be also expressed in
the basis of eigenmodes $\phi_n^{(m)}$ 
by $\psi_n(t) = \sum^{N}_{m=1} C_m(t)\, \phi_n^{(m)}$ (see below).
Here the coefficients $C_m(t)$ give the occupation probability
$\rho_m =  \langle\vert C_m(t) \vert^2\rangle$ where brackets denote
some long time or ensemble average (see below). The time evolution (\ref{eq1})
has two integrals of motion. They are: the (squared) norm
$\sum_n \vert \psi_n(t) \vert^2 = 1$ and total energy
$E = \sum_n [<\psi_n(t)|\hat{H}|\psi_n(t)> + 
(\beta/2) \vert \psi_n(t) \vert^4] $.
At $\beta=0$ the model (\ref{eq1}) can be viewed
as a quantum system or as a classical system of coupled linear
oscillators whose Hamiltonian in the basis of
oscillator eigenmodes is 
${\cal H} = \sum E^{\phantom *}_m C^*_m(t)\, C^{\phantom *}_m(t)$
with $C^{\phantom *}_m, C^*_m$ being a pair of conjugated variables;
$E_m$ plays the role of oscillator frequencies.

Due to the nonlinear term the eigenmodes are getting a nonlinear
frequency shift being 
$\delta \omega \sim \beta  \vert \psi_n \vert^2 \sim \beta /N$.
In \cite{chirikovyadfiz,dls1993,garcia} it was argued that
a developed chaos takes place when this shift $\delta \omega$ becomes 
comparable to a typical energy spacing between energies (or frequencies)
of the linear system $\Delta \omega \sim 1/N$. 
Thus  $\delta \omega > \Delta \omega$ implies chaos with the chaos border
 $\beta_c = const \sim 1$ being independent of system size $N$.
Thus above chaos border $\beta > \beta_c$
a moderate nonlinearity destroys KAM integrability leading to chaotic 
dynamics with a positive maximal Lyapunov exponent $\lambda$
and dynamical thermalization as it was shown in \cite{rmtprl}. 

The steady-state thermal distribution of probabilities $\rho_m$ 
has the standard form
corresponding to results of statistical mechanics \cite{landau}:
\begin{equation}
\rho_m = \rho_{EQ}(E_m)\equiv \frac{T}{E_m-\mu}
\label{eq2}
\end{equation}
corresponding to the equipartition of energies 
$\langle (E_m-\mu)|C_m|^2\rangle=(E_m-\mu)\rho_m=T$ where 
$T$ is the system temperature,
$\mu(T)$ is the chemical potential dependent on temperature.
These two parameters are determined from
the total norm\footnote{
For simplicity of notation we use the notation {\it norm}
for the quantity $\kappa$ even though it is 
a certain time averaged  norm  $|C_m(t)|^2$ of the quantum state with amplitudes 
$C_m(t)$. Actually, $\kappa$ is mathematically the 1-norm of the vector 
with coefficients $\rho_m$. Furthermore, we denote as {\it norm distribution} the 
dependence of $\rho_m$ on energies $E_m$ in particular for the 
cases where $\kappa\neq 1$. 
}
 $\kappa\equiv \sum_m \rho_m=1$ and the energy $E$, which are 
conserved integrals of motion, 
by the implicit equations $\kappa=\sum_m \rho_{EQ}(E_m) =1$ 
and  $\sum_m E_m \rho_{EQ}(E_m) =E$
(for $E$ we assume the case of a weak or moderate
nonlinearity which provides only a weak
contribution to the total energy).
The entropy $S$ of the system is given by
the usual relation \cite{landau}:
$S= - \sum_m \rho_m \ln \rho_m$
with the implicit  theoretical dependencies on temperature
$E(T)$, $S(T)$, $\mu(T)$ (see details in \cite{rmtprl}).
We note that a random matrix model similar to those
considered in \cite{rmtprl} and in (\ref{eq1})
was considered in \cite{prxshapiro}
but a detailed study of dynamical thermalization
was not presented there.

It is interesting to note that the dynamical thermal,
or Rayleigh-Jeans, distribution (\ref{eq2})
has been observed in optical multimode fibers 
\cite{fiber1,fiber2,fiber3,fiber4,fiber5}
(there a length $z$ along the fiber
corresponds to time variable discussed here).
At low temperatures $T$ the thermal distribution (\ref{eq2})
has maximal probabilities at low energy modes that was called 
self-cleaning in fibers.
At the same time we note that in all fiber experiments
\cite{fiber1,fiber2,fiber3,fiber4,fiber5}
the dynamics of rays in the linear system (at zero  nonlinear term) 
is always integrable usually corresponding to a case of two-dimensional 
oscillator (2D) potential with equal frequencies. 
The linear mode frequencies (or quantum energy levels)
of such a 2D oscillator are degenerate and formally the KAM theory is
not valid in such a situation. 
In particular, it was shown that for 3 oscillators with equal frequencies
about half of the phase-space is chaotic even at arbitrary small nonlinear
coupling \cite{chirikovyadfiz,mulansky2}. 
A similar situation appears also for equidistant
mode frequencies coupled by a nonlinear interaction in the FPU problem 
\cite{fpudls}. Therefore the situation of fiber experiments 
\cite{fiber1,fiber2,fiber3,fiber4,fiber5} does not directly correspond 
to the case of RMT mode frequencies or a case when a linear system belongs 
to the domain of quantum chaos \cite{bohigas,haake}
which has spectral properties being close to those of RMT.
We discuss relations between the NLIRM model (\ref{eq1}) with multimode 
fiber experiments below in more detail.

To model the KZ turbulence we generalize (\ref{eq1}) by 
adding terms for pumping and dissipation (absorption) 
at specific energy eigenmodes. 
For this it is more convenient to rewrite the time evolution equation 
(\ref{eq1}) by replacing $\psi_n$ with the amplitudes $C_m(t)$ obtained 
from the expansion of $\psi_n$ 
in the basis of the linear eigenmodes $\phi_n^{(m)}$. This provides 
the following generalized NLIRM model with pumping and dissipation~:
\begin{eqnarray}
\label{eq3}
i {{\partial C_m} \over {\partial {t}}}
&=& E_m C_m + i (\gamma_m - \sigma_m \vert C_m \vert^2)  C_m\\
\nonumber
&+& \beta \sum_{{m_1}{m_2}{m_3}}
V_{{m}{m_1}{m_2}{m_3}}
C_{m_1}C^*_{m_2}C_{m_3} \;\; .
%\label{eq3}
%\nonumber
\end{eqnarray}
The Hamiltonian case (\ref{eq1}) is a special case of 
(\ref{eq3}) with $\gamma_m=\sigma_m=0$ (taking into account 
the linear transformation $\psi_n\to C_m$). 
In (\ref{eq3}) the transitions between linear eigenmodes
appear only due to the nonlinear $\beta$-term
and the transition matrix elements are
$V_{{m}{m_1}{m_2}{m_3}}= \sum_{n}\phi_n^{(m)*}\phi_n^{(m_1)}
\phi_n^{(m_2)*}\phi_n^{(m_3)}\sim 1/N^{3/2}$ \cite{dls1993} due 
to the sum of $N$ random terms with typical size $N^{-2}$ 
since, according to RMT \cite{mehta}, $\phi_n^{(m)}\sim N^{-1/2}$. 
Furthermore, assuming ``random'' $C_m$ values of comparable size 
$C_m\sim C$ the $\beta$-term in (\ref{eq3}) has an overall size 
$\sim \beta C^3$ (sum of about $N^{3}$ random 
terms of typical size $V \sim N^{-3/2}$). 
%If the probability is distributed over 
%$N$ states of the system 
%then from the normalization
%condition we have $C_m \sim 1/N^{1/2}$
%and the transition rate to new 
%non-populated states in the basis $m$
%is $\Gamma \sim \beta^2 |C|^6 \sim \beta^2/(\Delta m)^3$.

In (\ref{eq3}), non-zero values of $\gamma_m>0$, $\sigma_m>0$ correspond 
to pumping modes or $\gamma_m<0$, $\sigma_m=0$ to dissipation modes. 
To obtain the energy flow of a direct cascade
from low to high energy modes $m$ we choose for pumping 
$\gamma_m=\gamma>0$ for the 4 lowest energy modes at $m=1,2,3,4$ 
with corresponding saturation coefficients
$\sigma_m = \sigma>0$ and for dissipation $\gamma_m=- \gamma<0, \sigma_m=0$
for the 4 highest energy modes with $m=N, N-1, N=2, N-3$. 
For all other $m$ values we chose $\gamma_m=\sigma_m=0$.
Here $\gamma$ and $\sigma$ are two parameters of our model 
and in most cases we choose $\gamma=\sigma=0.01$. 

To model an inverse cascade we also consider the case when
pumping is done at 4 $m$-values close to a certain 
pumping mode $m_0$ in mode space (with $m=m_0-2,\ldots,m_0+1$ 
and dissipation at system boundaries ($m=1,\ldots 4$ and $m=N-3,\ldots,1$). 

In this way we obtain a purely dynamical Random Matrix 
Model of KZ turbulence described by Eq. (\ref{eq3})
which we call RMM of KZT. This can be considered as a model of dynamical turbulence
without any couplings to an external thermal bath or external noise.

We mention, that 
in absence of the nonlinear coupling, i.e. if $\beta=0$, the amplitudes 
$C_m(t)$ decouple and (\ref{eq3}) allows for the analytical solution~:
\begin{equation}
\label{eq3solve}
C_m(t)=\frac{C_m(0)\,e^{-iE_mt}}{
\sqrt{D_m+(1-D_m)e^{-2\gamma_m t}}}\ ,\ 
D_m\equiv |C_m(0)|^2\frac{\sigma_m}{\gamma_m},
\end{equation}
which simplifies to $C_m(t)=C_m(0)\,e^{(-iE_m+\gamma_m)t}$ 
if $\sigma_m=0$. For the pumping case with 
$\gamma_m=\gamma>0$ and $\sigma_m=\sigma>0$ this solution provides 
$|C_m(t)|\to C_{\rm sat}\equiv \sqrt{\gamma/\sigma}$ for $t\to\infty$ 
with $C_{\rm sat}$ being the saturation value of the amplitudes. 
Initial small amplitudes $|C_m(0)|\ll C_{\rm sat}$ grow for short 
time scales as $|C_m(t)|\sim e^{\gamma t}$ and large initial 
amplitudes $|C_m(0)|\gg C_{\rm sat}$ decay as 
$|C_m(t)|\sim C_{\rm sat}/\sqrt{2\gamma t}$ for (very) short time scales 
and in both 
cases they saturate at $C_{\rm sat}$ with 
$\big||C_m(t)|-C_{\rm sat}\big|\sim e^{-2\gamma t}$ for 
longer time scales. 
For $\beta=0$ and dissipation modes, there is 
a simple exponential decay $|C_m(t)|\sim e^{-\gamma t}$. 
In all cases the phase $C_m(t)/|C_m(t)|= e^{-iE_mt}$ behaves as 
in the quantum or pure oscillator case.

It is interesting to note that in the limit 
of a strong $\beta$-term, or $E_m=0$, and $\gamma_m=\sigma_m=0$
Eq. (\ref{eq3}) is similar to
the random coupling model of turbulence
that can be considered as a classical
SYK model \cite{rosenhaus}.

\section{Theoretical KZ spectra  for RMM}
\label{sec3}

In the theory of KZ spectra, it is usually assumed
that the frequency $\omega$ spectrum of linear waves
is an algebraic function of the 
wave vector $k$ with $\omega(k) \propto k^\alpha$,
the four-wave interaction matrix elements are also algebraic 
functions of $k$ with an exponent $\chi$ ($V \propto k^\chi$)
and the system dimension is $d$. Then the stationary solution
of the direct cascade of energy 
flow from low to high energies also has an algebraic solution
with the 
density in $k$-space being \cite{zakharovbook} (see Eq.~(3.1.10a) there):
\begin{equation}
\rho_k \propto k^{-s_0}  \; , \; s_0=2\chi/3+d
\label{eq4}
\end{equation}
In our case for RMM
the wave vector $k$ corresponds to the eigenmode index $m$ and
we have $d=1$, $\chi=0$ (matrix elements $V$ in (\ref{eq3})
are independent of $m$) with $\alpha \approx 1$
(assuming a constant density of states for the center part of 
the semi-circle law we have approximately $E_m+1 \propto m$
with an energy shift counted from the lower energy border).
Hence, the theoretical steady-state density for RMM is
\begin{equation}
\rho_m \propto 1/(E_m+1)^{-s_0}  \; , \; s_0=1 \; .
\label{eq5}
\end{equation}

For the inverse cascade of norm flow in RMM we have from \cite{zakharovbook}
(see Eq.~(3.1.10b there):
\begin{equation}
\rho_m \propto (E_m+1)^{-x_0}  \; , \; x_0=2\chi/3+d - \alpha/3 = 2/3 \; .
\label{eq6}
\end{equation}

We perform for RMM of KZT a comparison of the above theoretical predictions
with the results of numerical simulations in next Section.

\section{Numerical results for RMM of KZT}
\label{sec4}

The numerical integration of the time evolution
described by equation (\ref{eq3}) 
is done in the same way as described in \cite{rmtprl}
using a fourth order integration method with a basic time step $\Delta t=0.1$. 
In absence of pumping and dissipation, the method conserves the 
symplectic symmetry and is also called ``symplectic integrator''. 
In particular in this case the total 
norm $\kappa = \sum_m |C_m(t)|^2$ is conserved up to numerical 
precision and the classical energy is conserved numerically 
at a level of $10^{-8}$. 
This integration method applies alternate small integration steps 
in the original lattice $n$-basis (only using the nonlinear $\beta$-term) 
followed by a small integration step of the other terms (obtained 
by putting $\beta=0$) 
in the linear eigenmode $m$-basis of the random matrix $\hat H$. Between 
two steps the linear transformation between $\psi_n\to C_m$ or 
its inverse transformation $C_m\to \psi_n$ is applied in order to switch 
forth and back from one to the other basis and vice versa. 
The succession of the 8 or 6 (using a certain symmetry optimization) 
individual integration time steps is carefully 
chosen as certain fractions 
of $\Delta t$ such that the global precision of the method is fourth order 
with a global error being $\sim (\Delta t)^4$ (see \cite{rmtprl} 
and references therein for details).

In the presence of pumping and dissipation terms the system described 
by (\ref{eq3}) is no longer symplectic, but the numerical method can 
still directly be applied to this case using the analytical solution 
(\ref{eq3solve}) for $\beta=0$ for the integration steps 
in $C_m$-representation and the overall precision is still $\sim (\Delta t)^4$.

These terms break norm and energy conservation
leading to turbulent energy and norm flows
between energy scales which are at the basis of
KZ spectra \cite{zakharovbook}. However, in-between
pumping and dissipation the dynamics (\ref{eq3})
remains Hamiltonian as it is assumed in KZ theory.

\begin{figure}[t]
\begin{center}
  \includegraphics[width=0.46\textwidth]{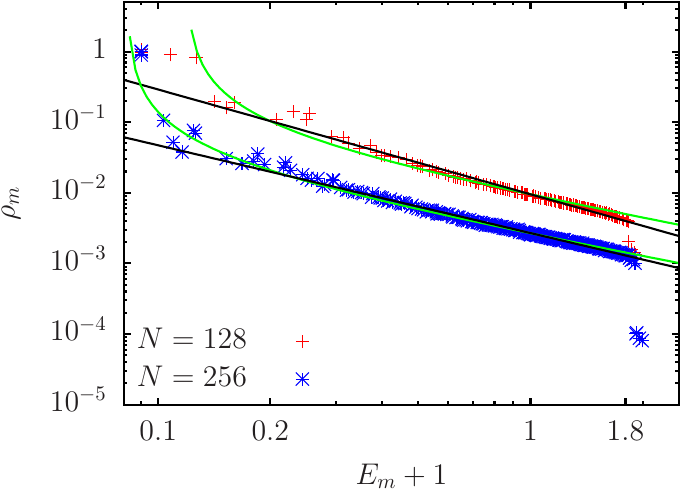}
\end{center}
\caption{\label{fig1}
Distribution $\rho_m$ versus $E_m+1$ in a double logarithmic representation 
for $N=128$ (red $+$) and $N=256$ (blue $*$), 
$\gamma=\sigma=0.01$, $\beta=1$, pumping 
at $m=1,\ldots, 4$ and absorption at $m=N-3,\ldots, N$. 
The values of $\rho_m$ have been obtained from the time average 
$\rho_m(t)=\langle |C_m(\tau)|^2\rangle$ for $t/2\le \tau\le t$ 
with $t=2^{22}$. 
The initial condition corresponds to uniform random values $C_m(0)$ 
for $m=1,\ldots, 8$ and $C_m(0)=0$ for $m>8$ with initial squared norm 
$\sum_m |C_m(0)|^2=0.01$. Data points with $E_m+1<0.09$ have been 
artificially moved to $E_m+1=0.09$. 
The black straight lines show the power law fit $\rho_m=A\,(1+E_m)^{-s_0}$ 
using the fit range $0.2\le 1+E_m\le 1.8$ 
with $A=(9.52\pm 0.10)\times 10^{-3}$, $s_0=1.48\pm 0.02$ (for $N=128$) 
and $A=(2.69\pm 0.01)\times 10^{-3}$, $s_0=1.24\pm 0.01$ (for $N=256$).
The average 
total norm is $\kappa=\sum_m \rho_m=6.25$ (for $N=128$) and 
$\kappa=5.31$ (for $N=256$). 
The green curves show the theoretical equipartition distribution 
$\rho_{EQ}(E_m)$ of Eq. (\ref{eq2}) obtained from a reduced spectrum 
$4<m<N-3$, excluding both pumping and dissipation modes with 
effective temperature $T=8.40\times 10^{-3}$ and chemical potential 
$\mu=-0.881$ (for $N=128$) and 
$T=2.45\times 10^{-3}$, $\mu=-0.918$ (for $N=256$; see text for details). 
The values of the reduced norm $\kappa_r$ and reduced energy $E_r$ 
used in Eq. (\ref{eq6a}) are 
$\kappa_r=2.489$, $E_r=-1.185$ (for $N=128$) 
and $\kappa_r=1.417$, $E_r=-0.6930$ (for $N=256$).
}
\end{figure}

\subsection{Direct cascade of KZT}

We start with the analysis of the direct energy cascade
from low to high energies where the pumping is done
at the lowest linear energy modes $m=1,\ldots,4$
and absorption (or dissipation) is done
at the highest energy modes $m=N-4, \ldots, N$.
We choose the initially populated modes
at $m=1,\ldots,8$ with random amplitudes $C_m(t=0)$
and a total norm $\kappa = \sum_m |C_m(0)|^2 = 0.01$.
A small value of $\kappa$ corresponds to an image of a calm
sea surface with very weak wind modeled by pumping. 
The results for system sizes $N=128, 256$, $\beta=1$, $\gamma=\sigma=0.01$ 
are presented in Fig.~\ref{fig1}.
They show that at large times
$t=2^{22}$ there is a convergence to a
stationary norm distribution $\rho_m$
which is approximately described by an algebraic
decay of the norm population $\rho_m$ with energy
$E_m+1$, counted from the spectrum border.
The decay exponent $s_0$, given by fit,
decreases with an increase of the system size
being $s_0=1.48; 1.24$ at $N=128, 256$.
The fit for the algebraic decay is done
for the range $0.2 \leq E_m+1 \leq 1.8$
to exclude the boundary effects of 
pumping and absorption.
Indeed, at low values of $m$ (or $E_m+1$)
the amplitudes $C_m(t)$ fluctuate around
the saturation value (for $\beta=0$) 
$C_m \approx C_{\rm sat}=\sqrt{\gamma/\sigma} =1$
corresponding to the regime when
$\gamma$-pumping is compensated by the nonlinear
$\sigma$-term. The total norm is $\kappa = 5.31$
showing that the largest fraction of norm
is located on the pumping modes $m=1,\ldots,4$
and a smaller norm fraction is transferred
by a turbulent flow to high energies.
At the boundary near $m=N$ the norm values $\rho_m$ drop  due to absorption.

\begin{figure}[t]
\begin{center}
  \includegraphics[width=0.46\textwidth]{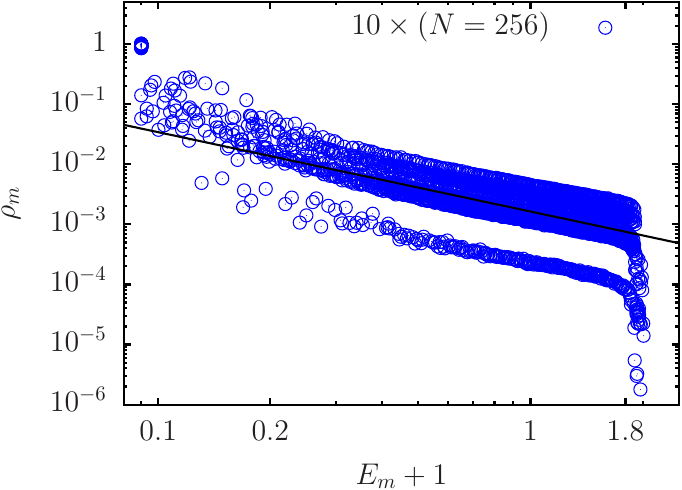}
\end{center}
\caption{\label{fig2}
As Fig.~\ref{fig1} for $N=256$ and 10 different 
random matrix realizations $\hat H$ (same other parameters 
and blue $\circ$ for data points). 
The straight line shows the power law fit $\rho_m=A\,(1+E_m)^{-s_0}$ 
using the fit range $0.2\le 1+E_m\le 1.8$ and all data points 
for all 10 random matrix realizations 
with 
$A=(1.64\pm 0.03)\times 10^{-3}$, $s_0=1.32\pm 0.03$.
The average  norm is in the interval $4.18\le \kappa =\sum_m \rho_m\le 6.57$ 
for the 10 random matrix realizations. 
}
\end{figure}

We point out 
that the steady-state distributions $\rho_m$,  shown in Fig.~\ref{fig1}
for a specific random realization of RMT matrix $H_{n,n'}$ in (\ref{eq1}),
are qualitatively the same for other random realizations of $H_{n,n'}$.
In Fig.~\ref{fig2}, we show steady-state distributions $\rho_m$
for 10 different realizations that have approximately the same shape
with an average algebraic exponent $s_0 = 1.32 \pm 0.03$.
The pumping region at $m =1, \ldots,4$ contains the main fraction
of the total norm $\kappa$ and depending on disorder realization of
$H_{n,n'}$ and $E_m$ there are considerable fluctuations of the 
numerical prefactor of the power law while the shape of 
the algebraic decay of $\rho_m$ at
higher energies remains independent of disorder.

\begin{figure}[t]
\begin{center}
  \includegraphics[width=0.46\textwidth]{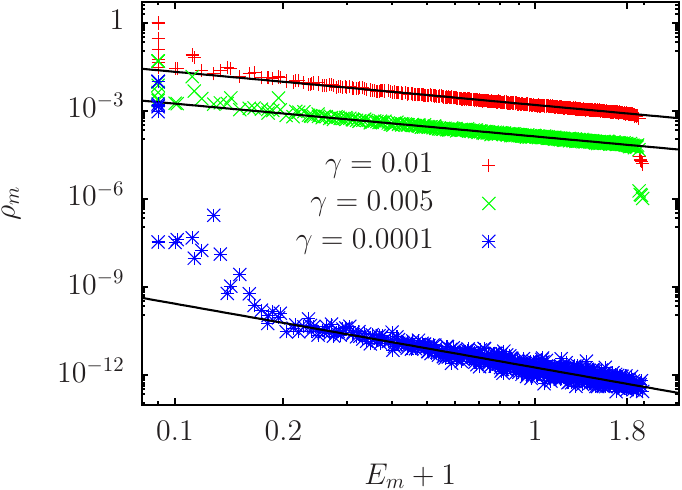}
\end{center}
\caption{\label{fig3}
As Fig.~\ref{fig1} but $N=512$ and modified 
values $\gamma=0.01$ (red $+$), $\gamma=0.005$ (green $\times$ and 
shifted down by a factor of 10 for better visibility) 
and $\gamma=0.0001$ (blue $*$). (Values for $\beta=1$, $\sigma=0.01$, 
$t=2^{22}$, modes for pumping/absorption 
and initial condition are unchanged with respect to Fig.~\ref{fig1}.)
The straight lines show the power law fit $\rho_m=A\,(1+E_m)^{-s_0}$ 
using the fit range $0.2\le 1+E_m\le 1.8$ 
with 
$A=(1.584\pm 0.003)\times 10^{-3}$, $s_0=1.130\pm 0.004$ (for $\gamma=0.01$), 
$A=(1.343\pm 0.004)\times 10^{-3}$, $s_0=1.117\pm 0.005$ (for $\gamma=0.005$) 
and 
$A=(1.82\pm 0.03)\times 10^{-12}$, $s_0=2.17\pm 0.03$ (for $\gamma=0.0001$) 
The average  total norm is $\kappa=5.99$ (for $\gamma=0.01$), 
$3.58$ (for $\gamma=0.005$) and 
$0.0455$ (for $\gamma=0.0001$).
}
\end{figure}

The fit values of $s_0$ in Fig.~\ref{fig1} and Fig.~\ref{fig2} 
are somewhat higher
than the theoretical value $s_0=1$ from (\ref{eq5}).
We attribute this to effects of finite system size $N$ and to 
deviations from the assumed constant density of states at the 
boundary of the semi-circle spectrum.
Indeed for the higher value $N=512$ shown in Fig.~\ref{fig3}, 
we obtain the exponent $s_0 = 1.130 \pm 0.004 $ being close to
the theoretical KZT value $s_0=1$ of Eq.~(\ref{eq5}) \cite{zakharovbook}.

One can also ask the question in how far the numerical results for 
Fig.~\ref{fig1} correspond to the statistical equipartition distribution 
(\ref{eq2}). Obviously, the numerical 
values of $\rho_m$ in Fig.~\ref{fig1} are significantly enhanced 
(reduced) for pumping (dissipation) modes $m$ if compared to $\rho_m$ for 
other close $m$ non-pumping (non-dissipation) modes and also from a 
theoretical view point it is not reasonable to expect that Eq. (\ref{eq2}) 
is valid for the full spectrum including pumping and dissipation modes. 
However, one may surmise that the long time ($t=2^{22}$) steady state 
distribution for the reduced spectrum with $4<m<N-4$, excluding 
pumping and dissipation modes, satisfies approximately the statistical 
equipartition distribution (\ref{eq2}). To verify this point we have 
determined $T$ and $\mu$ by the two implicit equations~:
\begin{equation}
\label{eq6a}
\kappa_r={\sum_m}' \rho_{EQ}(E_m)\quad,\quad
E_r={\sum_m}' E_m \rho_{EQ}(E_m)
\end{equation}
where the sums $\sum'$ are taken over the reduced spectrum $4<m<N-4$ 
and the reduced norm $\kappa_r$ and energy $E_r$ are obtained from the 
numerical values of $\rho_m$ at $t=2^{22}$ by~:
\begin{equation}
\label{eq6b}
\kappa_r\equiv{\sum_m}' \rho_m\quad,\quad
E_r\equiv {\sum_m}' E_m \rho_m\ .
\end{equation}
The green curves in Fig.~\ref{fig1} show $\rho_{EQ}(E_m)$ for 
the reduced spectrum of $N=128,256$, $\beta=1$, $\gamma=\sigma=0.01$ 
(the obtained values of $T,\mu,\kappa_r,E_r$ are given in the figure 
caption of Fig.~\ref{fig1}). These curves are rather close to 
the numerical data but for modes close to the energy 
boundaries there are some small but significant deviations and especially 
for energies in the interval $1<E_m<1.8$ the power law based on KZ theory 
(straight black lines) provide a slightly better fit than 
the classical equipartition distribution. 

Of course the distribution $\rho_{EQ}(E_m)$ in (\ref{eq2})
also can be viewed as an approximate algebraic decay $1/(E_m+1)$ at $E_m +1 \gg \mu$
with the algebraic exponent $s_0=1$ as in KZ theory (\ref{eq5}).
However, there is a fundamental physical
difference between the thermal steady-state (\ref{eq2})
and the turbulence steady-state  (\ref{eq5}):
there is no flow between energy scales in  (\ref{eq2})
while in  (\ref{eq5}) there is a turbulent flow from low to high energies
in (\ref{eq5}). Thus the empirical fits based on  (\ref{eq2})
for numerical data in Fig.~\ref{fig1} can be considered only as some additional
empirical descriptions without physical grounds behind.

We show in Fig.~\ref{fig3} the steady-state norm distribution $\rho_m$ 
for three values of pumping $\gamma =0.01; 0.005; 0.0001$ and 
the same value $\sigma=0.01$. 
For two largest $\gamma$ values the pumping is sufficiently strong
%(strong wind on a sea)
and the steady-state distribution
is rather close to the KZ theory. It is important to note
that this steady-state is practically independent of the initial
values of amplitudes $C_m(t=0)$ (for a given fixed realization of 
the RM $\hat H$). We have verified this point 
by choosing different random realizations of the initial condition 
(for the random initial values $C_m(0)$ for $m\le 8$ with 
$\kappa(t=0)=0.01$) and also localized initial conditions $C_{m_0}(0)=1$ 
for a specific mode value $m_0$ (e.g. with $m_0=1$ or $m_0=N/2$ etc.) 
and $C_m(0)=0$ for $m\neq m_0$ (typical initial condition 
chosen in \cite{rmtprl} with $\kappa(t=0)=1$). In all cases 
and for $N\le 512$, we obtain essentially the same steady-state 
distribution at $t=2^{22}$ with average relative differences 
$\langle \delta\rho_m/\rho_m\rangle$ of 
the order of $(1-2)\times 10^{-2}$ (for $N=128,256$) 
or $(5-10)\times 10^{-2}$ (for $N=512$). 
This shows that there a sigle global chaotic attractor
with describes a turbulent flow in the system.

According to Fig.~\ref{fig3} a sufficiently strong
pumping with $\gamma=0.01; 0.005$ leads to a turbulent flow and a steady-state
distribution $\rho_m$ in agreement with the KZ theory.
However, at small pumping with $\gamma=0.0001$ the distribution
$\rho_m$ has a very different shape with a fit exponent $s_0 = 2.17 \pm 0.03$.
In this case the total norm is reduced by a factor 100
with $\kappa=0.0455$,
corresponding to the estimate $\kappa \approx 4 \gamma/\sigma$
and a main norm fraction located on pumping modes $m=1,\ldots,4$ 
while $\rho_m\sim 10^{-12}-10^{-6}$ for other $m>4$. 
These results show that there is no turbulent flow at
a pumping strength being below a certain
chaos border that corresponds to the spirit of  KAM theory.
We stress that the notion of chaos border for emergence of the turbulence
flow is absent in the theory of KZ turbulence \cite{zakharovbook}.
We do not enter into a deep discussion on the value for the exponent 
$s_0 \approx 2$ in the non-turbulent regime,
but we note that due to RMT properties the matrix elements $V$ in (\ref{eq3})
have approximately the same random values and therefore 
based on simple perturbation theory with direct $V$-matrix element 
transitions one can estimate $C_m\sim 1/(E_m-E_1)$ 
and $\rho_m \sim 1/(E_m-E_1)^2$ (with $E_1\approx -1$)
corresponding to $s_0=2$ which is close to the numerical value $2.17$.

\begin{figure}[t]
\begin{center}
  \includegraphics[width=0.46\textwidth]{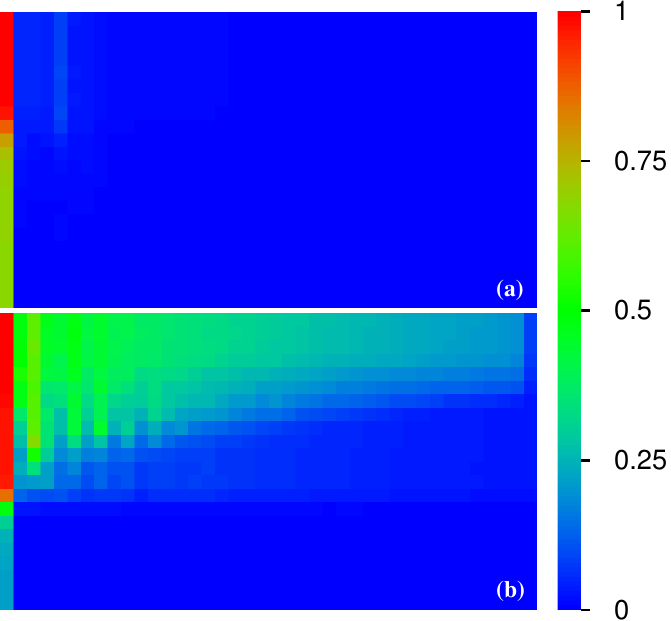}
\end{center}
\caption{\label{fig4}
Color density plot of $\langle\rho_m(t)\rangle$ 
in the plane $\log(E_m+1)$ ($x$-axis) 
and $\log_2(t)$ ($y$-axis) for the data of Fig.~\ref{fig2} 
with $\gamma=0.0001$ (a) and $\gamma=0.01$ (b) (same other parameters).
The values of $\rho_m(t)$ have been obtained from the time average 
$\rho_m(t)=\langle |C_m(\tau)|^2\rangle$ for $t/2\le \tau\le t$ 
with $t=2^{j}$ and the index $j=\log_2(t)=1,2,\ldots,22$ corresponding 
to the 22 vertical cells. The 40 horizontal cells correspond 
to 40 uniform cells in the interval $\log(0.09)\le \log(E_m+1)\le \log(2.1)$
with an additional average of $\rho_m(t)\to \langle\rho_m(t)\rangle$ 
for data points (with different $E_m$) in the same cell. 
Data points with $E_m+1<0.09$ have been artificially moved to $E_m+1=0.09$ 
and are taken into account in the average of the first column of cells. 
The values of the color bar correspond to 
$(\langle\rho_m(t)\rangle/\rho_{\rm max})^{1/4}$ 
with $\rho_{\rm max}=0.005058$ (a) and $\rho_{\rm max}=0.5051$ (b).
The top row of panel (a) corresponds to the bottom data points of 
Fig.~\ref{fig2} for $\gamma=0.0001$ and 
the top row of panel (b) corresponds to the top data points of 
Fig.~\ref{fig2} for $\gamma=0.01$ both for the last time value $t=2^{22}$. 
}
\end{figure}

The time development of turbulent flow from low to high energies
is shown in Fig.~\ref{fig4}. At low pumping below the chaos border 
with $\gamma=0.0001$ there is no flow to high energies 
while for sufficiently strong pumping
a turbulent flow to high energies emerges 
with a steady-state distribution in agreement with KZ theory.
The turbulent steady-state distribution approximately stabilizes 
at times $t \approx 2^{18}$. 

\begin{figure}[t]
\begin{center}
  \includegraphics[width=0.46\textwidth]{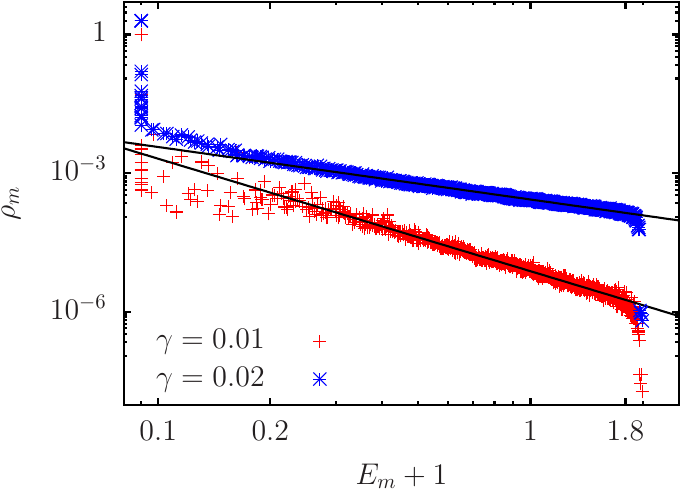}
\end{center}
\caption{\label{fig5}
As Fig.~\ref{fig1} for $N=1024$ and $\gamma=0.01$ (red $+$) 
or $\gamma=0.02$ (blue $*$) (same other parameters).
The straight lines show the power law fit $\rho_m=A\,(1+E_m)^{-s_0}$ 
using the fit range $0.2\le 1+E_m\le 1.8$ with 
$A=(7.64\pm 0.04)\times 10^{-6}$, $s_0=2.43\pm 0.01$ (for $\gamma=0.01$) 
and 
$A=(2.713\pm 0.005)\times 10^{-4}$, $s_0=1.134\pm 0.003$.
The total norm is $\kappa=4.15$ (for $\gamma=0.01$) and 
$9.07$ (for $\gamma=0.02$).
}
\end{figure}

Finally, in Fig.~\ref{fig5} we show results for the largest studied system size
$N=1024$. Here at $\gamma=0.01$ the algebraic decay exponent
$s_0 = 2.43 \pm 0.01$ is significantly different
from the theoretical value $s_0=1$ (\ref{eq5}). We attribute this
to the fact that the RMT density of states at lowest energies
drops significantly and thus with an increase of matrix size $N$
the spacing between lowest $E_m$ values becomes relatively larger
as compared to those in the middle of the energy band. Thus a higher
nonlinearity $\beta$ is needed to have a dynamical thermalization of
these initial eigenmodes. This effect has been discussed in \cite{rmtprl}
for Hamiltonian dynamics (\ref{eq1}).
Therefore a stronger pumping is required to be above a chaos border for
KZT. Indeed, the effective nonlinear parameter of the system is
$\beta \kappa \sim \beta \gamma/\sigma$ and 
an increase, of $\gamma$-pumping 
should drive the system to the KZ turbulent regime.
The results of Fig.~\ref{fig5} confirm that this is the case and at
$\gamma=0.02$ the exponent $s_0=1.134 \pm 0.003$
is in good agreement with the KZ theory.

\subsection{Regime of inverse cascade}

According to KZ theory \cite{zakharovbook}
there should also be an inverse turbulent cascade
of norm flow with the algebraic exponent $x_0=2/3$ (\ref{eq6}).
There have been experiments (see e.g. \cite{luka})
and numerical simulations (see e.g. \cite{shrira,nazar2023}) 
of realizations of such an inverse cascade.
We try to realize the regime of an inverse cascade
(for $\beta=1$, $\gamma=\sigma=0.01$, $N=512$) 
by performing $\gamma$-pumping in the same way as before 
but placing it either in the middle of the energy band $E_m$ (4 modes 
near $m_0=N/2$) or in the vicinity of maximal $E_m$ energies 
(4 modes near $m_0=N-40$). The absorption is done at both 
energy boundaries at 8 eigenmodes with 
$m=1,\ldots,4$ and $m=N-3, \ldots, N$ (using $\gamma_m=-\gamma<0$ 
and $\sigma_m=0$). 
The initial state is always composed of 8 eigenmodes placed around $m_0$
with random amplitudes and the total initial norm $\kappa(t=0) \approx 0.01$.

\begin{figure}[H]
\begin{center}
  \includegraphics[width=0.46\textwidth]{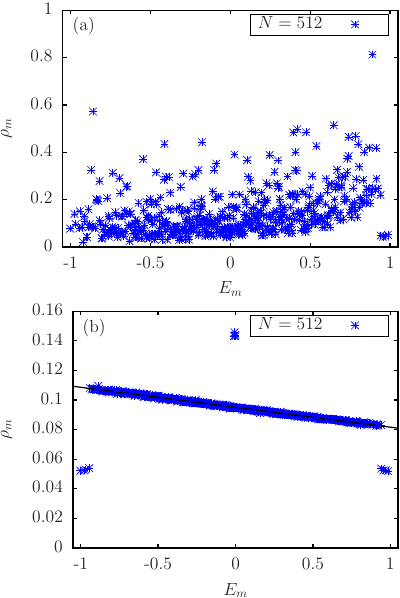}
\end{center}
\caption{\label{fig6}
Distribution $\rho_m$ versus $E_m$ in normal representation 
for $N=512$, $\beta=1$ and $\gamma=\sigma=0.01$.
The values of $\rho_m$ have been obtained from the time average 
$\rho_m(t)=\langle |C_m(\tau)|^2\rangle$ for $t/2\le \tau\le t$ 
with $t=2^{22}$. 
Both panels correspond 
to pumping at $m\approx m_0$ with $m_0=N-40=472$ (a) or $m_0=N/2=256$ (b), 
i.e. at $m=m_0-2,\ldots, m_0+1$ and absorption at both boundaries
$m=1,\ldots,4$ and $m=N-3,\ldots, N$. 
The initial condition corresponds to 8 uniform random values $C_m(0)$ 
at $m\approx m_0$, i.e. $m=m_0-4,\ldots, m_0+3$ 
and $C_m(0)=0$ for other $m$ with initial norm 
$\sum_m |C_m(0)|^2=0.01$. The final norm at $t=2^{22}$ 
is $\kappa=74.08$ (a) or $\kappa=48.52$ (b). 
The straight line (only for panel (b)) shows the linear fit 
$\rho_m=A-B\,E_m$ using non-pumping and non-absorption modes 
with $A=(9.51\pm 0.01)\times 10^{-2}$ and $B=(1.35\pm 0.02)\times 10^{-2}$. 
}
\end{figure}

\begin{figure}[t]
\begin{center}
  \includegraphics[width=0.46\textwidth]{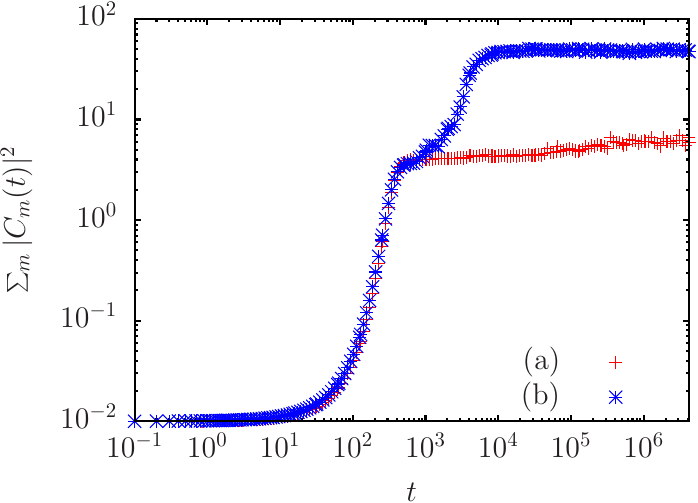}
\end{center}
\caption{\label{fig7}
Time dependence of the (non-averaged) 
norm $\sum_m |C_m(t)|^2$ at selected time values 
for $N=512$, $\beta=1$, $\gamma=\sigma=0.01$.
The pumping/absorption parameters are as in Fig.~\ref{fig2} for data 
points (a) (red $+$) or in (b) panel of Fig.~\ref{fig6} for data points (b) 
(blue $*$). 
Note that the total norm 
for times $\tau$ in the interval $2^{21}\le \tau\le 2^{22}$ 
is $\kappa=5.99$ (a) and $\kappa=48.52$ (b) 
(see also captions of Figs.~\ref{fig2} and \ref{fig5}). 
}
\end{figure}

Typical results are shown in Fig.~\ref{fig6}.
They clearly show an approximately homogeneous norm distribution
$\rho_m$ over all energies $E_m$; there are more fluctuations
for pumping at $m_0 = N-40$.
Clearly, there is no theoretical inverse cascade (\ref{eq6}).
We explain this fact as follows: a pumping at
high $m_0$ modes leads to a norm transfer at low 
$m$-modes as it is expected from KZ theory.
Thus an initially pumped norm $\kappa \approx 4\gamma/\sigma$
is transferred to low energy $m$-modes that leads to
a growth of the total norm accumulated in the system.
Indeed, the results of Fig.~\ref{fig7}
show that in the steady-state of the inverse cascade
the total norm $\kappa \approx 48$ is by a factor 10 larger
as compared to the case of the direct cascade with 
$\kappa \approx 6$. Thus for the inverse cascade case
the effective nonlinearity parameter
becomes rather large $\beta \kappa \approx 48$ 
therefore breaking the regime of weak or moderate
nonlinearity. It is possible that one can still
try to find a regime where the inverse cascade
is present in the RM model of KZT.
However, this would require to use very small
$\gamma$-pumping which may be 
below the chaos border. Furthermore, even at small $\gamma$ values
the total norm should significantly grow
in the limit of large system size $N$
where the weak turbulence approximation
is broken. For the case of a direct cascade
there is no such problem since 
$\rho_m$ is decreasing with growing energy $E_m+1$
and the total norm $\kappa$ with related
effective nonlinearity parameter $\beta \kappa$
remains bounded to moderate values (see Fig.~\ref{fig7})
and the system remains in a regime of weak turbulence
where KZ theory is valid.

Here we consider finite size systems with random matrix
couplings between  modes. Such a situation
corresponds to multimode fibers with a cross-section
generating chaotic ray dynamics corresponding
to a regime of quantum chaos (e.g. D-shape cross-section
as discussed in \cite{rmtprl}); such fibers may have
up to thousand of modes). Such systems are rather
different from quasi-infinite sizes used to study
weak turbulence of waves in large systems \cite{nazarenkobook,shrira,nazar2023}
(e.g. about billion modes in \cite{nazar2023}). It was shown that
at large sizes and a specific pumping the theoretical
inverse cascade can be realized \cite{nazar2023}. We do not exclude
that for some specific pumping and very large
random matrix sizes one can find regimes
with theoretical inverse cascade. But for moderate
sizes of about thousand, typical for fibers,
we tried various typical forms of pumping that were
always leading to a regime of strong nonlinearity.
Thus we conclude that the theoretical regime of inverse cascade
is hardly reachable for finite size systems
with random matrix interactions between modes,
e.g. chaotic fibers.
Also we think that for fibers it is rather difficult to
pump only very high modes.

\subsection{KZT and Anderson localization}

In the case of RMM described by Eq.~(\ref{eq1}) and Eq.~(\ref{eq3})
the matrix elements $V$ between linear eigenmodes
have random amplitudes of the same order between
all eigenmodes. It is interesting to consider a case
when such couplings have a local structure
with transitions only between modes in a certain finite
energy range of eigenmodes. As such a model
we consider the discrete Anderson nonlinear Schr\"odinger equation (DANSE)
in a static Stark field:

\begin{equation}
i {{\partial {\psi}_{n}} \over {\partial {t}}}
=(f n + \varepsilon_{n}){\psi}_{n}
+{\beta}|\psi_n|^2 \psi_{n}
 + ({\psi_{n+1}}+ {\psi_{n-1})}\; .
\label{eq7}
\end{equation}

Here $\varepsilon_n$ are random on-site energies
randomly and homogeneously distributed in the interval
$-W/2 \leq \varepsilon_n \leq W/2$, the hopping takes place only on
nearest sites with unit amplitude, $f$ describes a static Stark field.
At $f=\beta=0$ this system represents the one-dimensional Anderson model
with exponentially localized eigenmodes (see e.g. \cite{kramer,mirlin}).
The localization length at the middle of the energy band is
$\ell =96/W^2$. At moderate nonlinearity with $\beta \sim 1$ and $f=0$
this Anderson localization is destroyed and an unbounded
subdiffusive spreading of wave packet takes place \cite{dls2008,flach}. 
This spreading is preserved in presence of a moderate 
Stark field \cite{garcia2}.
In fact the spreading is subdiffusive such that 
the second moment grows as $\langle n^2\rangle \propto t^\nu$
with the numerical exponent  value $\nu \approx 0.3 - 0.4$
in 1D \cite{dls2008,flach,garcia2}. This value is smaller
than $\nu=0.5$ following from the random phase approximation \cite{basko,nazarmdpi}.
In \cite{rmtprl} it was shown that in such a system of finite size $N$
the dynamical thermalization takes place at a moderate 
nonlinearity $\beta \sim 1$ 
and a moderate Stark field $f$ with the steady-state thermal distribution
corresponding to energy equipartition (\ref{eq2}).

\begin{figure}[t]
\begin{center}
  \includegraphics[width=0.46\textwidth]{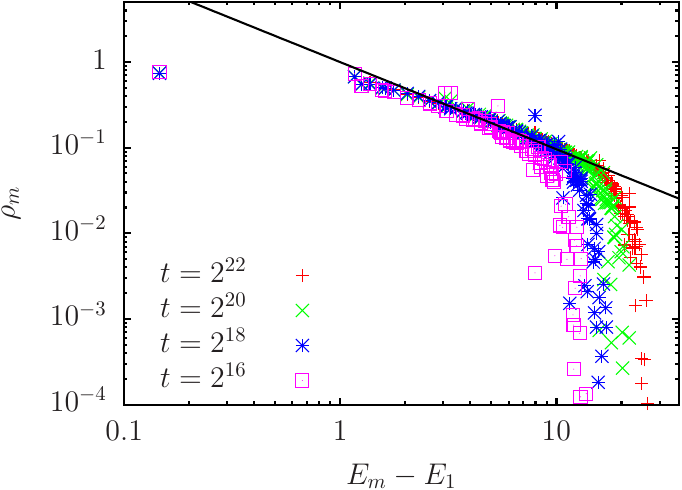}
\end{center}
\caption{\label{fig8}
$\rho_m$ versus $E_m-E_1$ in a double logarithmic representation 
for the DANSE model with diagonal shift parameter $f=0.125$, 
disorder strength $W=4$, $N=256$, $\gamma=\sigma=0.01$, $\beta=1$, 
and pumping/absorption/initial condition as in Fig. ~\ref{fig1}. 
The values of $\rho_m$ have been obtained from the time average 
$\rho_m(t)=\langle |C_m(\tau)|^2\rangle$ for $t/2\le \tau\le t$ 
with $t=2^{22}$ (red $+$), $t=2^{20}$ (green $\times$), $t=2^{18}$ (blue $*$) 
and $t=2^{16}$ (pink $\square$).
The straight line shows the power law fit $\rho_m=A\,(1+E_m)^{-s_0}$ 
using the data for $t=2^{22}$ and the fit range 
$0.01(E_{N}-E_1)\le (E_m-E_1)\le 0.5(E_{N}-E_1)$ 
with $A=1.00\pm 0.04$, $s_0=1.02\pm 0.02$ (and 
$E_1=-1.67,\,E_N=33.42,\,E_N-E_1=35.09$).
The total norm is
$\kappa=13.86$ (for $t=2^{16}$),
$17.16$ (for $t=2^{18}$),
$19.58$ (for $t=2^{20}$),
$21.19$ (for $t=2^{22}$). 
}
\end{figure}

\begin{figure}[t]
\begin{center}
  \includegraphics[width=0.46\textwidth]{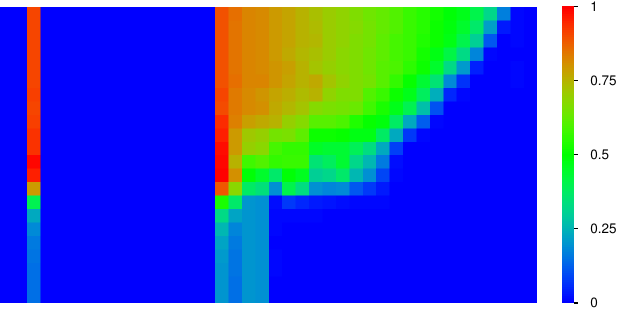}
\end{center}
\caption{\label{fig9}
Color density plot of $\langle\rho_m(t)\rangle$ 
in the plane $\log(E_m-E_1)$ ($x$-axis) 
and $\log_2(t)$ ($y$-axis) for the DANSE model using the 
data of Fig.~\ref{fig8} (similar style as in Fig.~\ref{fig4}). 
The values of $\rho_m(t)$ have been obtained from the time average 
$\rho_m(t)=\langle |C_m(\tau)|^2\rangle$ for $t/2\le \tau\le t$ 
with $t=2^{j}$ and the index $j=\log_2(t)=1,2,\ldots,22$ corresponding 
to the 22 vertical cells. The 40 horizontal cells correspond 
to 40 uniform cells in the interval 
$\log(0.1)\le \log(E_m-E_1)\le \log(1.1\times E_N)$ 
(same energy interval shown in Fig.~\ref{fig8})
with an additional average of $\rho_m(t)\to \langle\rho_m(t)\rangle$ 
for data points (with different $E_m$) in the same cell. 
The column 3 with first non-zero (non-blue) values corresponds to 
the data for $E_2$. The values of the color bar correspond to 
$(\langle\rho_m(t)\rangle/\rho_{\rm max})^{1/4}$ 
with $\rho_{\rm max}=1.072$. 
The top $1,3,5,7$-rows correspond to the data points of 
Fig.~\ref{fig8} for $t=2^{22},2^{20},2^{18},2^{16}$ respectively. 
}
\end{figure}

In the basis of linear eigenmodes of the 1d-Anderson Hamiltonian 
the coupling between modes 
appears only due to the $\beta$-nonlinearity. In this basis
the time evolution is described still by Eq.~(\ref{eq3}).
However, due to exponential localization of linear eigenmodes in (\ref{eq7})
the matrix elements $V$ are coupling $m$-modes only in a range
$\Delta m \sim \ell$ while outside of this range their amplitude
drops exponentially due to Anderson localization.

Due to the above properties it is natural to use
the DANSE model with Stark field for studies of KZT
by introducing pumping and absorption in the same way as for
RMM case described above. Another advantage of
the model (\ref{eq7}) is that due to the presence of the Stark field
the energy band range can be significantly increased as 
compared to the RMT case where it is fixed to $2$.

The results for KZ turbulence in model (\ref{eq7})
%%%are presented in Fig.~\ref{fig8}.
%%% Mentioning of Fig. 9
are presented in Figs.~\ref{fig8} and \ref{fig9}. 
%%%
Here as in RMM (for the direct cascade) the pumping
is done for the 4 lowest modes $m=1,\ldots,4$ at
$\gamma=\sigma=0.01$ and absorption is at the 
highest modes $m=N-4,\ldots,N$. 
The numerical integration is done in the same way as for RMM case.
We use parameters $\beta=1$, $W=4$ and $f=0.125$ so that at $N=256$
the energy band width is approximately 32 being by a factor 16
larger compared to RMM. At such parameters the localization length 
$\ell = 16$ is significantly smaller than the lattice size $N$.
The time evolution of the norm distribution
$\rho_m$ is shown in Fig.~\ref{fig8} at four discrete time values. 
%%% Discussion of new Fig. 9
The color density plot of Fig.~\ref{fig9} shows the same data as 
Fig.~\ref{fig8} but for all times $2\le t\le 2^{22}$ in the same style 
as Fig.~\ref{fig4} for the RMM. 
%%% 
It takes a rather long time to reach the absorption boundary
due to the slow subdiffusive spreading in the DANSE model.
However, with time the KZT profile (\ref{eq5}) stabilizes 
as well as the total norm growth.
In the steady-state we find
the algebraic decay exponent $s_0 = 1.02 \pm 0.02$
being practically equal to the theoretical value $s_0=1$.
We think that an increase of the energy band width
allows to obtain a more exact value of $s_0$.

We note that previously an interplay between
the KZ turbulence process and the Anderson localization
has been discussed in \cite{dlsturb1,dlsturb2}.
It was found that the KZ turbulent flow
to high energies can be stopped by
%%%the Anderson localization and KAN integrability.
%%% ``KAN'' is certainly an error ??? It should be KAM ???
the Anderson localization and KAM integrability.
However, in the models considered in  \cite{dlsturb1,dlsturb2}
the energy pumping  at low energy modes
was chosen to be unitary so that the total norm was conserved.
Due to this the nonlinear interaction was kept on a fixed level.
In the  case of model (\ref{eq7}) considered here the pumping
is changing the total norm so that the system itself relax
to a steady state with a self-determined total norm.
Therefore in such a case the Anderson localization cannot stop
the KZ turbulent flow which propagates to highest
available energies. However, as with the RMM case
(see Figs.~\ref{fig3},~\ref{fig4}),
a weak pumping in the DANSE model with Stark field (\ref{eq7}) 
keeps the system in the KAM integrable regime below 
the chaos border 
and there is no turbulent flow to high energies.

Finally, we note that for the case of the DANSE model (\ref{eq7}) with $f=0$
and with the pumping  done in the original $n$-basis we find that
the total norm is growing and the wave packet spreads over the whole
system size. Due to this  
norm growth of such a case is not of physical interest.

\section{Discussion}
\label{sec5}

In this work we introduced and studied a random matrix model
of the Kolmogorov-Zakharov turbulence. The model evolution is described
by purely dynamical equations of motion
without any external thermal bath or noise. As shown in \cite{rmtprl},
in absence of dynamical pumping and dissipation,
the Hamiltonian chaotic dynamics leads to a thermal equilibrium
distribution with energy equipartition over energies of linear modes.
Thus the dynamical evolution produces an emergence of  thermal
law description of statistical mechanics.
The introduction of energy and norm pumping at low energy modes 
and dissipation at high energy modes (also described by dynamical 
equations only) leads to a global chaotic attractor
which describes an energy flow from low to high energies
corresponding to the KZ theory concept \cite{zakharovbook}.
This direct turbulent flow is characterized 
by an algebraic decay of norm from low to high energies
with the exponent being close to the KZT theoretical value ($s_0=1$).
Thus the studied system can be viewed as a purely dynamical model
of KZ turbulence.
At the same time, we show that in this model  a turbulent flow appears only
at a pumping strength being above
a certain chaos border related to KAM integrability of motions
in the case of very weak nonlinearity.
We also show that the RMM case of KZT
is rather generic and even when linear eigenmodes
are exponentially localized in the lattice basis
due to Anderson localization we still have
the direct turbulent cascade well described
by KZ theory when pumping is above a certain chaos border.

However, in the RMM case we do not find a regime
of the inverse KZ cascade. We attribute this
to a strong norm growth driving the system to a strongly nonlinear regime
when the approach of weak turbulence and weak nonlinearity
is not applicable.

We note that the dynamical thermalization 
has been recently observed in the experiments with multimode fibers
(see e.g. \cite{fiber1,fiber2,fiber3,fiber4,fiber5})
and we assume that the RMM system discussed here 
for the KZ turbulence can be realized in fibers with
a cross-section of a chaotic 2D billiard. Such quantum billiards
have  many properties  of quantum chaos \cite{bohigas,haake}
being similar to the RMT case studied here.
As discussed in \cite{rmtprl}, 
we think that a most optimal billiard section of fiber
is a D-shape one (a circle with cut)
where the classical dynamics is know to be chaotic
(see e.g. \cite{cao}). We expect that in multimode fibers the KZT regime
can be realized by a continuous laser pumping of low energy modes
while the light at high energy modes will escape from a fiber
due to high angle collisions with fiber perimeter
thus leading to a stationary KZT flow from low to high energies.

 {\bf Acknowledgments:}
%\bigskip

This work has been partially supported through the grant
NANOX $N^o$ ANR-17-EURE-0009 in the framework of 
the Programme Investissements d'Avenir (project MTDINA).
This work was granted access to the HPC resources of 
CALMIP (Toulouse) under the allocation 2023-P0110.

%\clearpage

%\newpage
%\phantom{a}
%\newpage

%%%%%%%%%%%%%%%%%%%%%%%%%%%%%%%%%%%%%%%%%%%%%%%%%%%%%%%%%


\begin{thebibliography}{99}
\bibitem{kolm41} A.N.~Kolmogorov, 
       {\it The local structure of turbulence in an incompressible liquid
        for very large Reynolds numbers},
        Dokl. Akad. Nauk SSSR {\bf 30}, 299 (1941);
        {\it  Dissipation of energy in the locally isotropic turbulence},
        {\bf 32}, 19 (1941) [in Russian]
        (English trans. Proc. R. Soc. Ser. A {\bf 434}, 19 (1991);
        {\bf 434}, 15 (1991)).
\bibitem{obukhov} A.M.~Obukhov,
        {\it On energy distribution in the spectrum of a turbulent flow},
        Izv. AN SSSR Ser. Geogr. Geofiz., {\bf 5(4-5)}, 453 (1941) [in Russian]. 
\bibitem{filonenko} V.E.~Zakharov and N.N.~Filonenko,
       {\it Weak turbulence of capillary waves},
        J. Appl. Mech. Tech. Phys. {\bf 8 (5)}, 37 (1967).
\bibitem{zakharovbook} V.E.~Zakharov, V.~S.~L'vov and G.~Falkovich,
        {\it Kolmogorov spectra of turbulence},
        Springer-Verlag, Berlin (1992)
\bibitem{nazarenkobook} S.~Nazarenko, {\it Wave turbulence},
         Springer-Verlag, Berlin (2011).
\bibitem{luka} S.~Nazarenko and S.~Lukaschuk,
        {\it Wave turbulence on water surface},
        Annu. Rev. Condens. Matter Phys. {\bf 7}, 61 (2016).
\bibitem{galtier} S.~Galtier, {\it Physics of wave turbulence},
         Cambridge Univ. Press, Cambridhe UK (2023).
\bibitem{arnold} V.~Arnold, A.~Avez, {\it Ergodic problems of classical mechanics},
        Benjamin, N.Y. (1968).
\bibitem{sinai} I.~P.~Cornfeld, S.~V.~Fomin and Ya.~G.~Sinai,
        {\it Ergodic theory}, Springer-Verlag, N.Y. (1982).
\bibitem{chirikov} B.V.~Chirikov,
         {\it A universal instability of many-dimensional oscillator systems},
         Phys. Rep. {\bf 52}, 263 (1979).
\bibitem{lichtenberg} A.J.Lichtenberg, M.A.Lieberman, 
        {\it Regular and chaotic dynamics}, Springer, Berlin (1992).
\bibitem{rmtprl} K.M.Frahm and D.L.Shepelyansky,
         {\it Nonlinear perturbation of Random Matrix Theory},
         Phys. Rev. Lett. {\bf 131}, 077201 (2023).
\bibitem{wigner}  E.P.~Wigner, {\it Random matrices in physics},
        SIAM Review {\bf 9(1)}, 1 (1967).
\bibitem{mehta} M.L.~Mehta, {\it Random matrices},
         Elsvier, Amsterdam (2004).
 \bibitem{guhr} T.~Guhr,  A.~M\"uller-Groeling and H.A.~Weidenm\"uller,
          {\it Random Matrix Theories in quantum physics: common concepts},
          Phys.Rep. {\bf 299}, 189 (1998).
\bibitem{bohigas} O.~Bohigas, M.-J.~Giannoni and C.~Schmit,
         {\it Characterization of chaotic quantum spectra and
         universality of level fluctuation laws},
         Phys. Rev. Lett. {\bf 52}, 1 (1984).
\bibitem{haake} F.~Haake, {\it Quantum signatures of chaos},
         Springer, Berlin (2010).
\bibitem{landau} L.D.~Landau and E.M.~Lifshitz, 
        {\it Statistical physics}, 
        Wiley, New York (1976).        
\bibitem{prxshapiro} A.~Ramos, L.~Fernandez-Alcazar, T.~Kottos and B.~Shapiro,
        {\it Optical phase transitions in photonic networks: a spin-system formulation}
        Phys. Rev. X {\bf 10}, 031024 (2020).
\bibitem{fpu} E.Fermi, J.Pasta and S.Ulam,
        {\it Studies of non linear problems},
         Los Alamos Report LA-1940 (1955); published later in 
         E.Fermi {\it Collected papers}, E.Serge (Ed.) {\bf 2}, 491,
         Univ. Chicago Press, Chicago IL (1965);
          see also historical overview in T.Dauxois
          {\it Fermi, Pasta, Ulam and a mysterious lady},
           Phys. Today {\bf 61(1)}, 55 (2008).
           % https://doi.org/10.1063/1.2835154
\bibitem{greene} C.~S.~Gardner, J.~M.~Greene, M.~D.~Kruskal
         and R.~M.~Miura, {\it Method for solving
         the Korteweg - de Vries equation},
         Phys. Rev. Lett. {\bf 19}, 1095 (1967).
\bibitem{zakharov} V.E.~Zakharov and A.B.~Shabat,
          {\it Interaction between solitons in a stable medium},
          Sov. Phys. JETP {\bf 37(5)}, 823 (1973).
\bibitem{toda} M.~Toda, {\it Studies of a non-linear lattice},
            Phys. Reports {\bf 18(1)}, 1 (1975).
\bibitem{landau6} L.D.~Landau and E.M.~Lifshitz, 
        {\it Fluid mechanics}, 
        Pergamon Press, New York (1989). 
\bibitem{lapteva}  T.V.~Laptyeva, A.A.~Tikhomirov, O.I.~Kanakov,
          and  M.V.~Ivanchenko,
            {\it Anderson attractors in active arrays},
            Sci. Reports {\bf 5}, 13263 (2015).
\bibitem{rlaser} G.~Rollin, J.~Lages and D.L.~Shepelyansky,
        {\it Random lasing from Anderson attractors},
          Phys. Rev. A {\bf 105}, 053506 (2022). 
\bibitem{chirikovyadfiz} B.V.~Chirikov and D.L.~Shepelyanskii,
         {\it Dynamics of some homogeneous models of
          classical Yang-Mills fields},
           Sov. J. Nucl. Phys. {\bf 36(6)}, 908 (1982). 
\bibitem{dls1993} D.L.~Shepelyansky, 
         {\it  Delocalization of quantum chaos by weak nonlinearity},
         Phys. Rev. Lett. {\bf 70}, 1787 (1993).
\bibitem{garcia} I.Garcia-Mata and D.L.Shepelyansky,
        {\it Delocalization induced by nonlinearity in systems with disorder},
        Phys. Rev. E {\bf 79}, 026205 (2009).
\bibitem{fiber1} K.~Baudin , A.~Fusaro, K.~Krupa, J.~Garnier, S.~Rica, G.~Millot, and A.~Picozzi,
        {\it Classical Rayleigh-Jeans condensation of light waves: observation and
           thermodynamic characterization}, Phys. Rev. Lett. {\bf 125}, 244101 (2020).
\bibitem{fiber2} N.~Berti, K.~Baudin, A.~Fusaro, G.~Millot, A.~Picozzi, and J.~Garnier,
        {\it Interplay of thermalization and strong disorder: wave turbulence theory,
          numerical simulations, and experiments in multimode optical fibers},
        Phys. Rev. Lett. {\bf 129}, 063901 (2022).
\bibitem{fiber3} E.V.~Podivilov, F.~Mangini,  O.S.~Sidelnikov,  M.~Ferraro,  M.~Gervaziev, 
        D.S.~Kharenko, M.~Zitelli, M.P.~Fedoruk, S.A.~Babin, and S.~Wabnitz,
        {\it Thermalization of orbital angular momentum beams in multimode optical fibers},
          Phys. Rev. Lett. {\bf 128}, 243901 (2022).
\bibitem{fiber4} F.~Mangini, M.Gervaziev,  M.~Ferraro,  D.S.~Kharenko,
        M.~Zitelli,  Y.~Sun, V.~Couderc,  E.V.~Podivilov, S.A.~Babin, and S.Wabnitz,
        {\it Statistical mechanics of beam self-cleaning in GRIN multimode optical fibers},
         Optics Express {\bf 30(7)}, 10850 (2022)
\bibitem{fiber5}  K.~Baudi, J.~Garnier, A.~Fusaro, N.~Berti, C.~Michel, K.~Krupa, G.~Millot
        and A.~Picozzi, {\it  Observation of light thermalization to
        negative-temperature Rayleigh-Jeans
        equilibrium states in multimode optical fibers},
        Phys. Rev. Lett. {\bf 130}, 063801 (2023).
\bibitem{mulansky2}  M.~Mulansky, K.~Ahnert, A.~Pikovsky and D.L.~Shepelyansky,
         {\it Strong and weak chaos in weakly nonintegrable many-body Hamiltonian systems},
           J. Stat. Phys. {\bf 145}, 1256 (2011).
\bibitem{fpudls} D.L.~Shepelyansky,
         {\it Low energy chaos in the Fermi-Pasta-Ulam problem},
           Nonlinearity {\bf 10}, 1331 (1997).    
\bibitem{rosenhaus} X.-Y.~HU and V.~Rosemhaus,
         {\it The random coupling model of turbulence as a classical SYK model},
          arXiv:2303.03421 [hep-th] (2023).
\bibitem{shrira} S.Y.~Annenkov and V.I.~Shrira,
         {\it   Direct numerical simulation of downshift and
         inverse cascade for water wave turbulence},        
         Phys. Rev. Lett. {\bf 96}, 204501 (2006).
\bibitem{nazar2023} Y.~Zhu, B.~Semisalov, G.~Krstulovic and S.~Nazarenko,
         {\it Direct and inverse cascades in turbulent Bose-Einstein condensates},
         Phys. Rev. Lett. {\bf 130}, 133001 (2023).
\bibitem{kramer} B.~Kramer and A.~MacKinnon,
         {\it  Localization: theory and experiment},
         Rep. Prog. Phys. {\bf 56}, 1469 (1993).
\bibitem{mirlin} F.~Evers and A.D.~Mirlin,
         {\it Anderson transitions}, Rev. Mod. Phys. {\bf 80}, 1355 (2008).
\bibitem{dls2008} A.S.~Pikovsky, and D.L.~Shepelyansky, 
         {\it Destruction of Anderson localization by a weak nonlinearity}, 
         Phys. Rev. Lett. {\bf 100}, 094101 (2008).
\bibitem{flach} T.V.~Lapteva, M.I.~Ivanchenko, and S.~Flach, 
         {\it Nonlinear lattice waves in heterogeneous media}, 
         J. Phys. A: Math. Theor. {\bf 47}, 493001 (2014).
\bibitem{garcia2} I.Garcia-Mata and D.L.Shepelyansky,
         {\it Nonlinear delocalization on disordered Stark ladder},
           Eur. Phys. J. B {\bf 71}, 121 (2009).
\bibitem{basko} D.~Basko,
         {\it   Kinetic theory of nonlinear diffusion in a weakly disordered nonlinear
         Schrodinger chain in the regime of homogeneous chaos},         
         Phys. Rev. E {\bf 89}, 022921 (2014).
\bibitem{nazarmdpi}  S.~Nazarenko, A.~Soffer and M.-B.~Tran,
         {\it On the wave turbulence theory for the nonlinear
           Schrodinger equation with random potentials},
         Entropy {\bf 21}, 823 (2019).
\bibitem{dlsturb1}  D.L.~Shepelyansky,
         {\it Kolmogorov turbulence, Anderson localization and KAM integrability},
          Eur. Phys. J. B {\bf 85}, 199 (2012).
\bibitem{dlsturb2} L.Ermann, E.Vergini and D.L.Shepelyansky,
         {\it Kolmogorov turbulence defeated by Anderson localization
           for a Bose-Einstein condensate in a Sinai-oscillator trap},
         Phys. Rev. Lett. {\bf 119}, 054103 (2017).
\bibitem{cao} K.~Kim, S.~Bittner, Y.~Jin, Y.~Zeng, Q.J.~Wang, and H.~Cao,
           {\it Impact of cavity geometry on microlaser dynamics},
         Phys. Rev. Lett. {\bf 131}, 153801 (2023).
         
\end{thebibliography}
\end{document}